# Negative differential conductance effect and electrical anisotropy of 2D $ZrB_2$ monolayers

Yipeng An[1,2,5], Jutao Jiao[1], Yusheng Hou[2], Hui Wang[2], Ruqian Wu[2,5], Chengyan Liu[2], Xuenian Chen[3], Tianxing Wang[1] and Kun Wang[4]

1. College of Physics and Materials Science & International United Henan Key Laboratory of Boron Chemistry and Advanced Energy Materials, Henan Normal University, Xinxiang 453007, China
2. Department of Physics and Astronomy, University of California, Irvine, California 92697, USA
3. School of Chemistry and Chemical Engineering & International United Henan Key Laboratory of Boron Chemistry and Advanced Energy Materials, Henan Normal University, Xinxiang 453007, China
4. Department of Mechanical Engineering, University of Michigan, Ann Arbor, Michigan 48109, USA

E-mail: ypan@htu.edu.cn (Y An), wur@uci.edu (R Wu)

---

[5] Author to whom any correspondence should be addressed.

**Abstract**

Two-dimensional (2D) metal-diboride $ZrB_2$ monolayers was predicted theoretically as a stable new electronic material [A. Lopez-Bezanilla, *Phys. Rev. Mater.*, 2018, **2**, 011002 (R)]. Here, we investigate its electronic transport properties along the zigzag (z-$ZrB_2$) and armchair (a-$ZrB_2$) directions, using the density functional theory and non-equilibrium Green's function methods. Under low biases, the 2D $ZrB_2$ shows a similar electrical transport along zigzag and armchair directions as electric current propagates mostly *via* the metallic Zr-Zr bonds. However, it shows an electrical anistropy under high biases, and its $I-V$ curves along zigzag and armchair directions diverge as the bias voltage is higher than 1.4 V, as more directional B-B transmission channels are opened. Importantly, both z-$ZrB_2$ and a-$ZrB_2$ show a pronounced negative differential conductance (NDC) effect and hence they can be promising for the use in NDC-based nanodevices.

# 1. Introduction

Since borophene was successfully synthesized in 2015, boron and boron-based monolayer materials have attracted a large wave of research interest in recent years[1-6]. Several different boron monolayers have been predicted by means of the first-principles calculations[3, 4], and they are promising for a variety of applications. To date, boron monolayers such as out-of-plane buckling borophenes[1], $\beta_{12}$ and $\chi_3$ types of boron sheets[2] have been synthesized on the silver substrate. To develop a truly freestanding two-dimensional (2D) borophene, which is unstable according to theoretical calculations[7, 8], several growth strategies have been proposed, including surface hydrogenation[7] and binding with metal atoms[9-16]. As is known, the structural instability of the graphene-like boron monolayer is due to its electron deficiency. Naturally, one may combine metal atoms that have a suitable size and chemical activity with boron to achieve geometric stability. To this end, a few metal-boron nanostructures have been predicted, including metal-diborides (such as $MgB_2$, $TiB_2$, $FeB_2$, and $ZrB_2$)[9-12] and other metal-borides (such as $CrB_4$, $TiB_4$, $FeB_6$, and $MnB_6$) [13-16], and some of them have been successfully prepared in experiments[17]. For instance, $MgB_2$ monolayers have recently been fabricated on the Mg(0001) substrate *via* molecular beam epitaxy[17], following the theoretical prediction of Tang *et al* by density functional theory (DFT) calculations[9].

Very recently, a new metal-diboride monolayer, $ZrB_2$, was predicted by the first-principles calculations[12]. Interestingly, ZrB$_2$ is structurally stable and exhibits a unique electronic structure with two Dirac cones out of the high-symmetry lines in the irreducible Brillouin zone (BZ)[12]. The ZrB$_2$ nanosheets are expected to be prepared in experiments like the MgB$_2$ monolayers by the means of molecular beam epitaxy[17]. For the use of this material in nanodevices, it is crucial to investigate 2D ZrB$_2$ monolayers more holistically, especial for its electronic transport properties. A few important aspects need to be examined: (i) does the 2D ZrB$_2$ monolayer have any peculiar current−voltage (*I*−*V*) behaviors? (ii) how strong is its electrical anisotropy? (iii) is there any unique feature for device applications? (iv) whether defects are important?

In this report, we systematically study the electronic transport properties of 2D ZrB$_2$ monolayer (see figure 1(a)) through the first-principles calculations. We find that the 2D ZrB$_2$ monolayer exhibits similar electrical transport behavior along the zigzag and armchair directions under low biases, as its *I*−*V* curves are almost overlapped. However, it becomes anisotropic when the bias goes beyond 1.4 V. The current mainly propagates *via* hopping along the Zr-Zr bonds under low biases, whereas additional B-B transmission channels are opened with a high bias. Importantly, ZrB$_2$ shows a pronounced negative differential conductance (NDC) effect, which may sustain even with the presence of defects. Therefore, ZrB$_2$ monolayer can be a promising candidate for the NDC-based nanodevices.

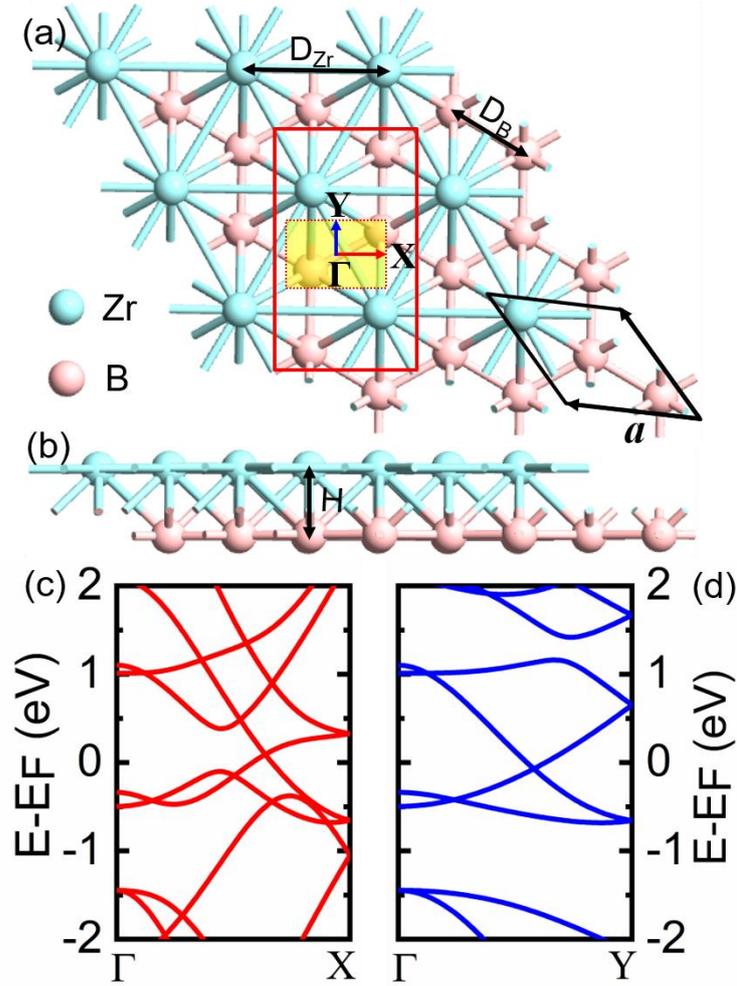

**Figure 1.** Top (a) and side (b) views of the two-dimensional ZrB$_2$ monolayer. The black/red box in (a) refers to its hexagonal/simple orthorhombic unit cell. The first Brillouin zone of simple orthorhombic unit cell is embedded in its box. Band structures of the ZrB$_2$ unit cell along the two *k*-lines $\Gamma$–X (c) and $\Gamma$–Y (d). The Fermi level is set to zero.

## 2. Computational methods

The electronic structures and transport properties of the 2D ZrB$_2$ monolayer are determined by using the density functional theory and the nonequilibrium Green's function approaches as implemented in the Atomistix Toolkit (ATK) code[18-21]. The electron exchange and correlation effect is described within the Perdew-Burke-Ernzerhof (PBE) scheme of the generalized gradient approximation (GGA)[22, 23]. For all Zr and B atoms, their core electrons are described by the optimized Norm-Conserving Vanderbilt (ONCV) pseudo-potentials, and wave functions of valence states are expanded as linear combinations of atomic orbitals (LCAO), at the level of SG15[24] pseudo-potentials and basis sets. Note that the SG15 datasets of ONCV pseudo-potentials are fully relativistic and can provide comparable results with the all-electron approach. The atomic structures are fully relaxed until the residual force on each atom becomes less than 1 meVÅ$^{-1}$ and the energy tolerance is below $10^{-6}$ eV, respectively. We use an energy cutoff 150 Ry for the basis expansion and a 1×150×150 Monkhorst-Pack *k*-points grid to sample the 2D Brillouin zone (BZ) for the structural optimization and band calculations. For the electronic transport calculations, we use a 1×9×150 *k*-points grid to sample the Brillouin zone of the left and right electrodes, to achieve a balance between the accuracy and cost.

## 3. Results and Discussions

Figure 1 (a) and (b) presents the top and side views of 2D ZrB$_2$ monolayer. The B atoms are arranged in a honeycomb lattice and Zr atoms are placed above the center of B hexagons. After optimization, the B-B bond length (D$_B$) is 1.82 Å, about the same as that in the bulk ZrB$_2$ (1.83 Å)[25], and the height (H) of Zr atoms is 1.53 Å. The lattice parameter *a* is 3.16 Å (equals to the Zr-Zr distance, D$_{Zr}$), in good agreement with its bulk phase obtained in recent experiments[26]. The band structures of its hexagonal unit cell is shown in figure S1, also in consistent with results in the literature[12].

The 2D ZrB$_2$ monolayer has two different surfaces, one only has Zr atoms (Zr surface) and the other is graphene-like borophene (borophene surface). It may exhibit a metallic characteristic like the out-of-plane buckling borophenes[1], $\beta_{12}$ and $\chi_3$ types of boron sheet[2]. Therefore, it is an interesting topic to know which surface would dominate the electronic transport in different bias ranges. The borophene surface has zigzag and armchair rows (denoted as z-$ZrB_2$ and a-$ZrB_2$, respectively), it is hence also interesting to investigate how strong its electrical anisotropy is. Indeed, the $ZrB_2$ exhibits some distinctive behaviors in its band structures along the zigzag ($\varGamma$–$X$) and armchair ($\varGamma$–$Y$) directions in figure 1(c) and (d). While both have two bands crossing the Fermi level ($E_F$), there are obviously more bands around the Fermi level along the $\varGamma$–$X$ line.

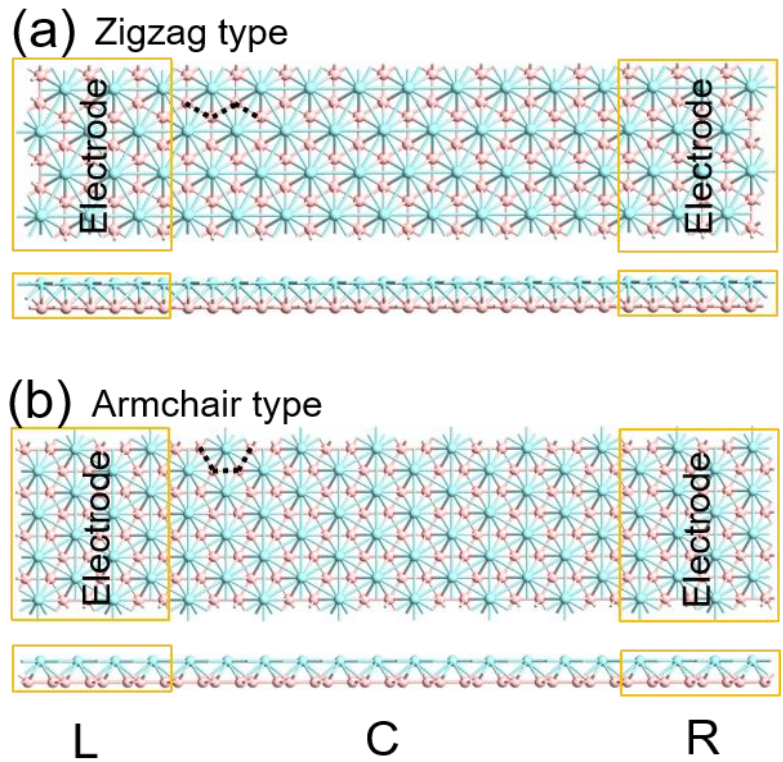

**Figure 2.** The two-terminal structures of zigzag(a) and armchair type (b) $ZrB_2$ monolayer. The top/bottom part in (a) and (b) refers to their top/side view. L/R refers to the left/right electrode, and C represents the central scattering region.

We construct a nanodevice structure based on the 2D ZrB$_2$ monolayer (see figure 2), including the two types along the zigzag and armchair directions, and study their electronic transport properties in detail. In each direction, it can be regarded as a two-terminal structure, i.e., z-ZrB$_2$ type and a-ZrB$_2$, as shown in figure 2(a) and (b). The 2D supercells have a periodicity perpendicular to the direction of current between the left (L) and right (R) electrodes. The third direction is out of the plane, along which the slabs are separated by a 30 Å vacuum. Both the L and R electrodes are semi-infinite in length along the transport direction and are described by a large supercell. When a bias $V_b$ is applied across the L and R electrodes, their energies are shifted accordingly. A positive bias gives rise to an electric current from the L electrode to the R electrode, and vice versa. In the present work, the current $I$ through the z-ZrB$_2$ (figure 2(a)) and a-ZrB$_2$ (figure 2(b)) two- terminal structures is obtained by using the Landauer–Büttiker formula[27]

$$I(V_b) = \frac{2e}{h}\int_{-\infty}^{\infty} T(E,V_b)[f_L(E-\mu_L) - f_R(E-\mu_R)]dE, \quad (1)$$

where $T(E,V_b)$ is the bias-dependent transmission coefficient, calculated from the Green's functions; $f_{L/R}$ are the Fermi-Dirac distribution functions of the left/right electrodes; $\mu_L$ (= $E_F - eV_b/2$) and $\mu_R$ (= $E_F + eV_b/2$) are the electrochemical potentials of the left and right electrodes, respectively. For more details, one can see previous descriptions for this method in the literature[18-20].

Figure 3(a) shows the current−voltage curves of the z-ZrB$_2$ and a-ZrB$_2$ two-terminal structures, biased in the voltage range [0 to 2.0] V. It is interesting to find that their $I-V$ curves overlap well under a low bias less than 1.4 V. However, when the applied bias is beyond this critical point, the 2D ZrB$_2$ monolayer shows a strong anisotropic current (see figure 3(a)) like the 2D borophene and borophane [28, 29]. Specially, the ZrB$_2$ has higher conductance along the zigzag direction than along the armchair one under high biases (> 1.4 V). The ratio of current anisotropy $\eta=I_z/I_a$ ($I_z$ and $I_a$ refer to currents of z-ZrB$_2$ and a-ZrB$_2$, respectively) is about 1.7 at 2.0 V, equal to that of borophane[28] and larger than that of borophene[29]. Moreover, both *I−V* curves show a pronounced negative differential conductance effect[30] in the bias range [1.0 to 1.4] V for the z-ZrB$_2$ and [0.9 to 1.6] V for the a-ZrB$_2$ (see figure 3(b), the curve of *dI/dV* vs. the voltage). Their differential conductances *dG* (= *dI/dV*) go oscillates, with minimum values of −103.58 $\mu$S for z-ZrB$_2$ at V=1.2 V bias and −98.25 $\mu$S for a-ZrB$_2$ at 1.4 V. The maximum conductivity of ZrB$_2$ is about 1.6×10$^5$ S/m, less than that of Al (3.8×10$^7$ S/m), Au (4.5×10$^7$ S/m), and Ag (6.2×10$^7$ S/m)[31].

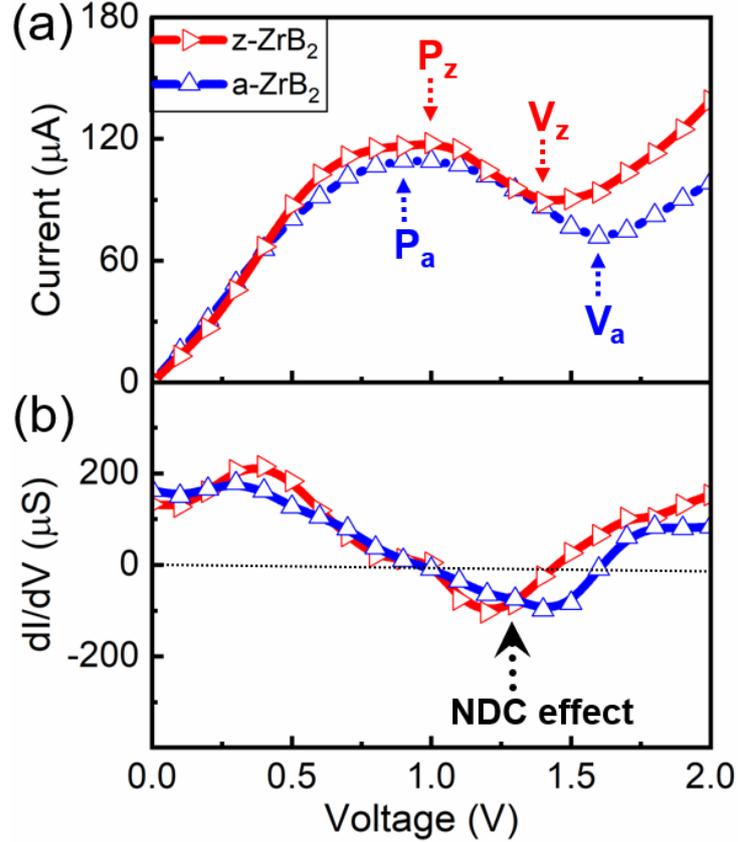

**Figure 3.** *I−V* (a) and *dG−V* curves of the zigzag and armchair type ZrB$_2$. P$_i$/V$_i$ in (a) refers to the peak/valley value of electric current.

The NDC effect is the main feature of *I−V* curves of resonant tunneling diodes initially proposed by Esaki and his co-workers[32, 33]. It has been extensively studied not only for its counterintuitive nature, but also due to its potential applications in nanoelectronics[34-37], such as high-frequency oscillators, multipliers, mixers, logic, and analog-to-digital converters[38-40]. This effect is characterized by two key factors that can be obviously dependent on the materials and vary substantially in experiments. The first factor is the so-called peak-to-valley ratio (PVR) between the maximal (peak) and minimal (valley) currents. The second one is the NDC voltage where the current reaches its maximum. In general, it is desired to have large PVR but low NDC voltage for minimizing the power consumption. For 2D $ZrB_2$ monolayer, its PVR is ~1.33:1 (i.e., $P_z$:$V_z$) for z-$ZrB_2$, and 1.52:1 (i.e., $P_a$:$V_a$) for a-$ZrB_2$. The latter is slightly larger than that of a thiol-terminated Ru(∥) bis-terpyridine molecular junction (~1.47:1)[41]. On the other hand, the NDC voltage of the $ZrB_2$ is only ~1.0 V, smaller than those of other materials[42]. Therefore, the 2D $ZrB_2$ monolayer can be a good candidate for the use in NDC-based nanodevices.

Differing from the resonant tunneling mechanism proposed in Esaki's work[32, 33], the electronic transport properties of 2D monolayers mostly depend on their band structures. Namely, electron transmission contains dominating contributions from intra- and inter-band transitions around the Fermi level. To unveil the physical origin of the NDC effect of $ZrB_2$, we calculate and analyze its transmission spectra and projected band structures along the zigzag and armchair directions under nonzero biases. Figures 4(a)-(c) show results of the z-$ZrB_2$ under a bias of 1.0 V (corresponding to its current peak), 1.4 V (corresponding to its current valley), and 2.0 V. The bands of the left and right electrodes shift down and up (see figures 4(a)-(c)) as the bias increases, respectively.

Both the overlap and component of their energy bands determine the electron transmission, which increases gradually (see figure S2) and gives large coefficients (such as ~4.69 at the Fermi level) when the bias reaches to 1.0 V (see figure 4(a)). The integral over the BW (by equation 1) equals to the local maximum current ($P_z$, see figure 3(a)). Noted that under a low bias (such as 1.0 V), electron transmission mainly results from Zr orbitals according to the band structures (see figure 4(a)). As the bias window increases, the transmission coefficients drop (see figure S2) obviously. It is mainly due to that, the weight from the B/Zr states increases/decreases in the right electrode while not in the left electrodes (see figure 4(b)), although the total band overlap numbers change little. The competition between these two factors (i.e., the increase of bias window and decrease of transmission coefficients) leads to a current reduction and the formation of a valley in the *I*−*V* curve (i.e., $V_z$ in figure 3(a)). Under a high bias (such as 2.0 V), for both the left and right electrodes, more bands of B states enter the bias window and open more transmission channels (see figure 4(c)), which leads to a wide transmission peak and an uptrend of the *I*−*V* curve. To understand this NDC effect, we also analyze the maps of the transmission coefficients as a function of energy and wavevector K along the periodical direction (see figure 4(e)-(f)). Increasing the bias causes the extension of the bottleneck effect [43] in the transmission near the positive bias window. The bottleneck of transmission coefficient gradually enters the bias window as increasing the bias from 1.0 to 1.4 V (see figure 4(d) and (e)), which causes the current reducing and the NDC effect appearing. However, this effect disappears after 1.4 V due to the increasing contributions from the electrons with negative energy which have larger transmission probability (see figure 4(f)). The same NDC mechanism is also applicable to the a-$ZrB_2$, and its bias-dependent transmission spectra and electrode bands are given in figure S3. With a high bias (such as 2.0 V), a large gap is appeared in the band overlap of the left and right electrodes (see figure S3 (c)). It results a transmission gap at the corresponding energy region (from 0.05 to 0.25 eV) and depressed *I−V* curve compared to the z-ZrB$_2$. Thus, a rigid electronic anisotropy is appeared for ZrB$_2$.

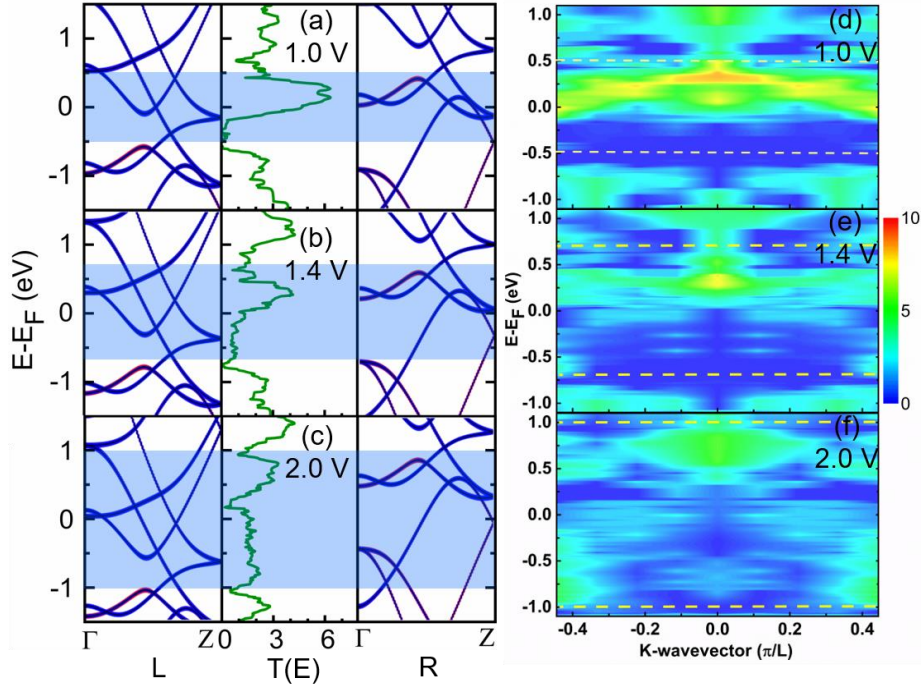

**Figure 4.** Transmission spectra and projected band structures for the left and right electrodes of the z-ZrB$_2$ under the biases of 1.0 (a), 1.4 (b), and 2.0 (c) V, respectively. The weights of Zr and B atoms are indicated in blue and red, respectively. (e)-(f) show the maps of transmission coefficients as a function of energy and wavevector K under various biases. L in (f) is the lattice length of z-ZrB$_2$ along the periodic direction. The Fermi energy of band structures is shifted to zero. The shadows in (a)-(c) and dashed lines in (e)-(f) denote the bias window.

The electron transmission pathway is an analysis option that splits the transmission coefficient into local bond contributions $T_{ij}$[44]. The pathways across the boundary between two parts, A and B, give the total transmission coefficient

$$T(E) = \sum_{i \in A, j \in B} T_{ij}(E). \quad (2)$$

In general, there are two types of local current pathways: via chemical bonds (i.e., bond current) or via electron hopping (i.e., hopping current) between atoms[45]. Our results indicate that Zr-Zr bond currents play the leading role for both z-ZrB$_2$ and a-ZrB$_2$ (see figure 5(a) and (b)) under low biases. Although there is a difference between Zr-Zr bonds between z-ZrB$_2$ and a-ZrB$_2$ (namely, the bond current of z-ZrB$_2$ is parallel to the transport direction, whereas it makes an angel away from the horizontal axis for a-ZrB$_2$.), it gives rise to insignificant influence on their electron transmissions (see figure 4(a) and S3(a)) due to the metallic feature of the Zr-Zr bonds. As more directional B-B transmission channels are opened (see figure 5(c)) under high biases (> 1.4 V), anisotropic conduction in the B-layer prevails. However, transmission pathways for a-ZrB$_2$ is very few at the Fermi level under the bias of 2.0 V (see figure 5(d)), due to the presence of a large gap in the band overlap and transmission spectrum (see figure 3(c)).

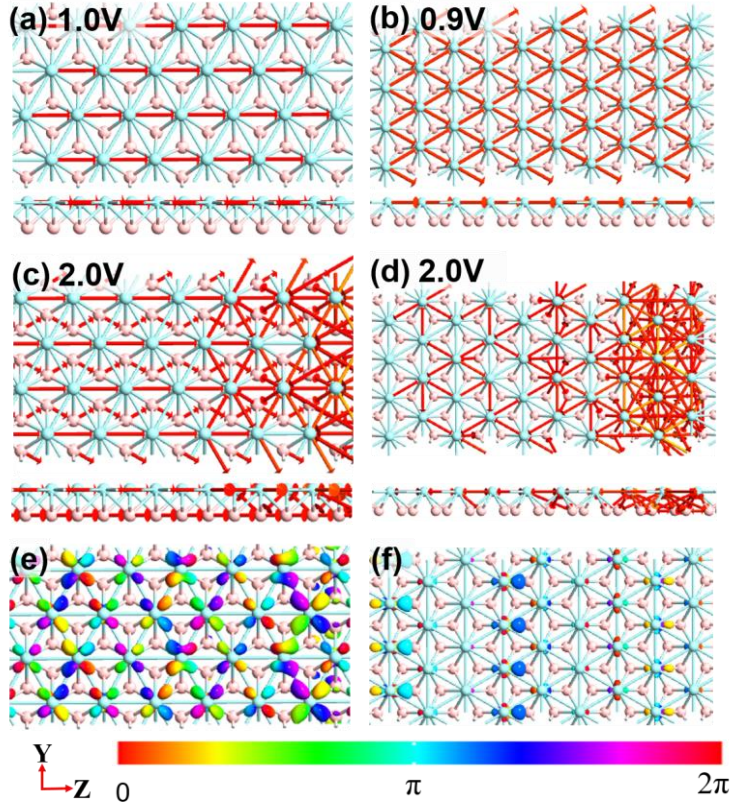

**Figure 5.** The transmission pathways of the z-ZrB$_2$ at 1.0 (a) and 2.0 V (c), and the a-ZrB$_2$ at 0.9 (b) and 2.0 V (d). The arrows point in the directions of electric current flow. The primary transmission eigenstates TE-I at the Fermi level for the z-ZrB$_2$ at 1.0 V (e) and a-ZrB$_2$ at 0.9 V (f).

As shown by the transmission spectrum in figure 4(a), the transmission coefficient of the z-ZrB$_2$ is ~4.69 at the E$_F$. This means that there exist at least 5 degenerated transmission channels (because this is a single-electron transmission and the transmission coefficient of each channel should be no more than 1). Its primary transmission eigenstate (labeled as z-TE-I) is depicted in figure 5(e), and the other four secondary TEs are shown in figure S4. We may see that the z-TE-I stems mostly from the $4d_{y^2-z^2}$ orbitals of Zr atoms. For a-ZrB$_2$, the transmission coefficient is ~3.94 at the E$_F$ (see figure S3(a)) and it thus has four transmission channels contributed by a primary transmission eigenstate (i.e., a-TE-I). As displayed in figure 5(f), this state is mainly composed of the 4$d_{yz}$ orbitals of Zr atoms. Feature of the other three secondary TEs are shown in figure S5, respectively. Noted that the transmission eigenstates, including its degeneracy and spatial composition, depend on the bias. The change of band weight and overlapping region between the left and right electrodes may give different transmission eigenstates and conduction channels.

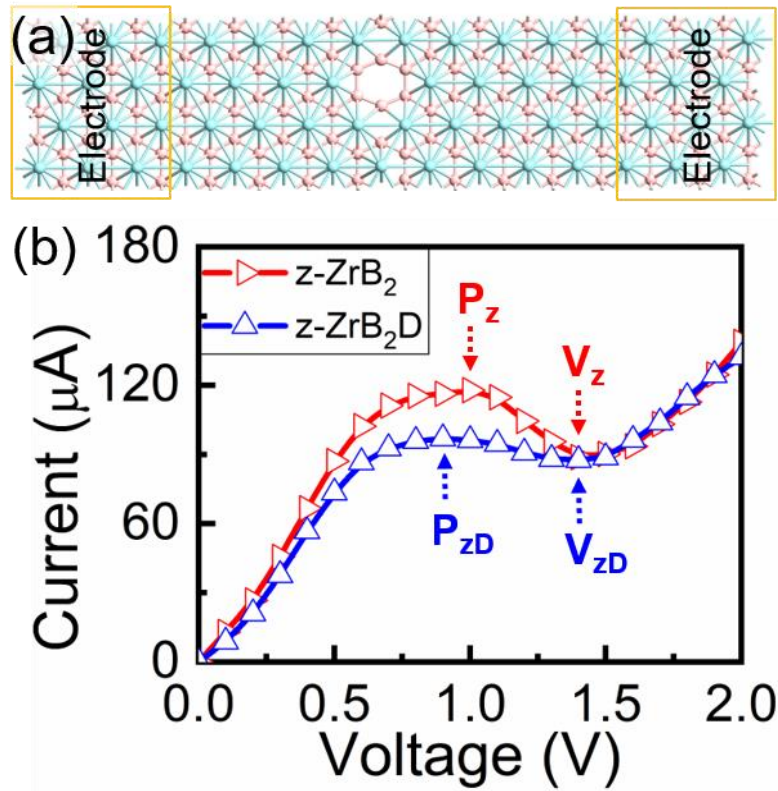

**Figure 6.** The two-terminal structure (a) and $I-V$ curve (b) of the z-ZrB$_2$ with Zr vacancy.

In practical experiments, defects (such as a vacancy) in samples are generally inevitable. It is crucial to demonstrate whether the NDC effect can be changed if ZrB$_2$ samples has defects. Here, we investigate the electronic transport properties of z-ZrB$_2$ with a Zr vacancy (labeled as z-ZrB$_2$D), in the structure shown in figure 6(a). The $I-V$ curve of this z-ZrB$_2$D structure is given in figure 6(b). Interestingly, the conductance decreases only a little in a broad range of bias. The NDC effect can still be observed but the PVR changes from 1.33:1($P_z$:$V_z$) to 1.11:1 ($P_{zD}$:$V_{zD}$). Overall, the presence of Zr vacancies is expected to degrade the NDC performance of $ZrB_2$ devices. Therefore, one needs to improve the quality of 2D $ZrB_2$ monolayer materials for applications.

## 4. Conclusions

In conclusion, by using the first-principles density functional theory and the non-equilibrium Green's function approaches, we systematically study the electronic transport properties of 2D $ZrB_2$ monolayers along the zigzag and armchair directions. Our results demonstrate that $ZrB_2$ shows a similar (anisotropic) electrical transport along the two orthogonal directions under low (high) biases, primarily depending on if current goes through the metallic Zr-Zr or covalent B-B channels. Interestingly, the $I-V$ curves display pronounced NDC effect around a bias voltage beyond 1.0 V. This feature may sustain even with the presence of Zr vacancies, and hence $ZrB_2$ monolayers can be promising for the use in NDC-based nanodevices.

**Acknowledgements**

The work at the University of California at Irvine was supported by the US DOE-BES under Grant DE-FG02-05ER46237. The work at Henan Normal University was supported by the National Natural Science Foundation of China (Grant Nos. 11774079 and U1704136), the CSC (Grant No. 201708410368), the Natural Science Foundation of Henan Province (Grant No. 162300410171), the young backbone teacher training program of Henan province's higher education, the Science Foundation for the Excellent Youth Scholars of Henan Normal University (Grant No. 2016YQ05), and the High-Performance Computing Centre of Henan Normal University.

**References**

[1] Mannix A J, Zhou X-F, Kiraly B, Wood J D, Alducin D, Myers B D, Liu X, Fisher B L, Santiago U, Guest J R, Yacaman M J, Ponce A, Oganov A R, Hersam M C and Guisinger N P 2015 *Science* **350** 1513

[2] Feng B, Zhang J, Zhong Q, Li W, Li S, Li H, Cheng P, Meng S, Chen L and Wu K 2016 *Nature Chem.* **8** 563

[3] Wu X, Dai J, Zhao Y, Zhuo Z, Yang J and Zeng X C 2012 *ACS Nano* **6** 7443

[4] Zhou X-F, Dong X, Oganov A R, Zhu Q, Tian Y and Wang H-T 2014 *Phys. Rev. Lett.* **112** 085502

[5] Zhang Z, Mannix A J, Hu Z, Kiraly B, Guisinger N P, Hersam M C and Yakobson B I 2016 *Nano Lett.* **16** 6622

[6] Mannix A J, Zhang Z, Guisinger N P, Yakobson B I and Hersam M C 2018 *Nat. Nanotech.* **13** 444

[7] Xu L-C, Du A and Kou L 2016 *Phys. Chem. Chem. Phys.* **18** 27284

[8] Jena N K, Araujo R B, Shukla V and Ahuja R 2017 *ACS Appl. Mater. Interfaces* **9** 16148

[9] Tang H and Ismail-Beigi S 2009 *Phys. Rev. B* **80** 134113

[10] Zhang L, Wang Z, Du S, Gao H-J and Liu F 2014 *Phys. Rev. B* **90** 161402

[11] Zhang H, Li Y, Hou J, Du A and Chen Z 2016 *Nano Lett.* **16** 6124

[12] Lopez-Bezanilla A 2018 *Phys. Rev. Mater.* **2** 011002

[13] Lopez-Bezanilla A 2018 *2D Mater.* **5** 035041

[14] Qu X, Yang J, Wang Y, Lv J, Chen Z and Ma Y 2017 *Nanoscale* **9** 17983

[15] Li J, Fan X, Wei Y, Liu J, Guo J, Li X, Wang V, Liang Y and Chen G 2016 *J. Mater. Chem. C* **4** 10866

[16] Li J, Wei Y, Fan X, Wang H, Song Y, Chen G, Liang Y, Wang V and Kawazoe Y

2016 *J. Mater. Chem. C* **4** 9613

[17]   Bekaert J, Bignardi L, Aperis A, van Abswoude P, Mattevi C, Gorovikov S, Petaccia L, Goldoni A, Partoens B, Oppeneer P M, Peeters F M, Milošević M V, Rudolf P and Cepek C 2017 *Sci. Rep.* **7** 14458

[18]   Taylor J, Guo H and Wang J 2001 *Phys. Rev. B* **63** 121104

[19]   Brandbyge M, Mozos J-L, Ordejón P, Taylor J and Stokbro K 2002 *Phys. Rev. B* **65** 165401

[20]   Soler J M, Artacho E, Gale J D, García A, Junquera J, Ordejón P and Sánchez-Portal D 2002 *J. Phys.: Condensed Matter.* **14** 2745

[21]   *Atomistix ToolKit. Available: http://quantumwise.com/.*

[22]   Perdew J P, Chevary J A, Vosko S H, Jackson K A, Pederson M R, Singh D J and Fiolhais C 1992 *Phys. Rev. B* **46** 6671

[23]   Perdew J P, Burke K and Ernzerhof M 1996 *Phys. Rev. Lett.* **77** 3865

[24]   Schlipf M and Gygi F 2015 *Comput. Phys. Commun.* **196** 36

[25]   Post B, Glaser F W and Moskowitz D 1954 *Acta Metall.* **2** 20

[26]   Baris M, Simsek T, Simsek T, Ozcan S and Kalkan B 2018 *Adv. Powder Technol.* **29** 2440

[27]   Büttiker M, Imry Y, Landauer R and Pinhas S 1985 *Phys. Rev. B* **31** 6207

[28]   Padilha J E, Miwa R H and Fazzio A 2016 *Phys. Chem. Chem. Phys.* **18** 25491

[29]   Shukla V, Grigoriev A, Jena N K and Ahuja R 2018 *Phys. Chem. Chem. Phys.* **20** 22952

[30]   Xu B and Dubi Y 2015 *J. Phys.: Condens. Matter.* **27** 263202

[31]   *http://periodictable.com/Properties/A/ElectricalConductivity.an.html.*

[32]   Chang L L, Esaki L and Tsu R 1974 *Appl. Phys. Lett.* **24** 593

[33]   Esaki L 1958 *Phys. Rev.* **109** 603

[34]   Fan Z Q, Zhang Z H, Deng X Q, Tang G P, Yang C H, Sun L and Zhu H L 2016 *Carbon* **98** 179

[35]   An Y P, Zhang M J, Wu D P, Fu Z M and Wang K 2016 *J. Mater. Chem. C* **4** 10962

[36]   Zhang R Q, Li Z Y and Yang J L 2017 *J. Phys. Chem. Lett.* **8** 4347

[37] Li Z, Zheng J X, Ni Z Y, Quhe R G, Wang Y Y, Gao Z X and Lu J 2013 *Nanoscale* **5** 6999

[38] Brown E R, Soderstrom J R, Parker C D, Mahoney L J, Molvar K M and McGill T C 1991 *Appl. Phys. Lett.* **58** 2291

[39] Mathews R H, Sage J P, Sollner T, Calawa S D, Chen C L, Mahoney L J, Maki P A and Molvar K M 1999 *Proc. IEEE* **87** 596

[40] Broekaert T P E, Brar B, van der Wagt J P A, Seabaugh A C, Morris F J, Moise T S, Beam E A and Frazier G A 1998 *IEEE J. Solid-State Circuits* **33** 1342

[41] Zhou J F, Samanta S, Guo C L, Locklin J and Xu B Q 2013 *Nanoscale* **5** 5715

[42] Grobis M, Wachowiak A, Yamachika R and Crommie M F 2005 *Appl. Phys. Lett.* **86** 204102 204102

[43] Nguyen M C, Nguyen V H, Nguyen H-V and Dollfus P 2015 *J. Appl. Phys.* **118** 234306

[44] Solomon G C, Herrmann C, Hansen T, Mujica V and Ratner M A 2010 *Nature Chem.* **2** 223

[45] Castro Neto A H, Guinea F, Peres N M R, Novoselov K S and Geim A K 2009 *Rev. Mod. Phys.* **81** 109